\documentclass[aps,twocolumn,amsmath,letterpaper,floatfix]{revtex4}
\usepackage{amssymb}
\usepackage{graphicx}
\usepackage{array}
\usepackage{hhline}
\usepackage{longtable}

\begin{document}

\title{Infinitely Robust Order and Local Order-Parameter Tulips

in Apollonian Networks with Quenched Disorder}
\author{C. Nadir Kaplan,$^{1, 2}$ Michael Hinczewski,$^{3, 4}$ and A. Nihat Berker$^{1,5,6}$}
\affiliation{$^1$Department of Physics, Ko\c{c} University, Sar\i
yer 34450, Istanbul, Turkey,} \affiliation{$^2$Martin Fisher School
of Physics, Brandeis University, Waltham, Massachusetts 02454,
U.S.A.,}\affiliation{$^3$Feza G\"ursey Research Institute, T\"UBITAK
- Bosphorus University, \c{C}engelk\"oy 34684, Istanbul, Turkey,}
\affiliation{$^4$Department of Physics, Technical University of
Munich, 85748 Garching, Germany,} \affiliation{$^5$Faculty of
Engineering and Natural Sciences, Sabanc\i~University, Orhanl\i ,
Tuzla 34956, Istanbul, Turkey,} \affiliation{$^6$Department of
Physics, Massachusetts Institute of Technology, Cambridge,
Massachusetts 02139, U.S.A.}

\begin{abstract}
For a variety of quenched random spin systems on an Apollonian
network, including ferromagnetic and antiferromagnetic bond
percolation and the Ising spin glass, we find the persistence of
ordered phases up to infinite temperature over the entire range of
disorder.  We develop a renormalization-group technique that yields
highly detailed information, including the exact distributions of
local magnetizations and local spin-glass order parameters, which
turn out to exhibit, as function of temperature, complex and
distinctive tulip patterns.
\end{abstract}
\maketitle
\def\s{\rule{0in}{0.28in}}

\begin{figure*}
\includegraphics*[scale=1]{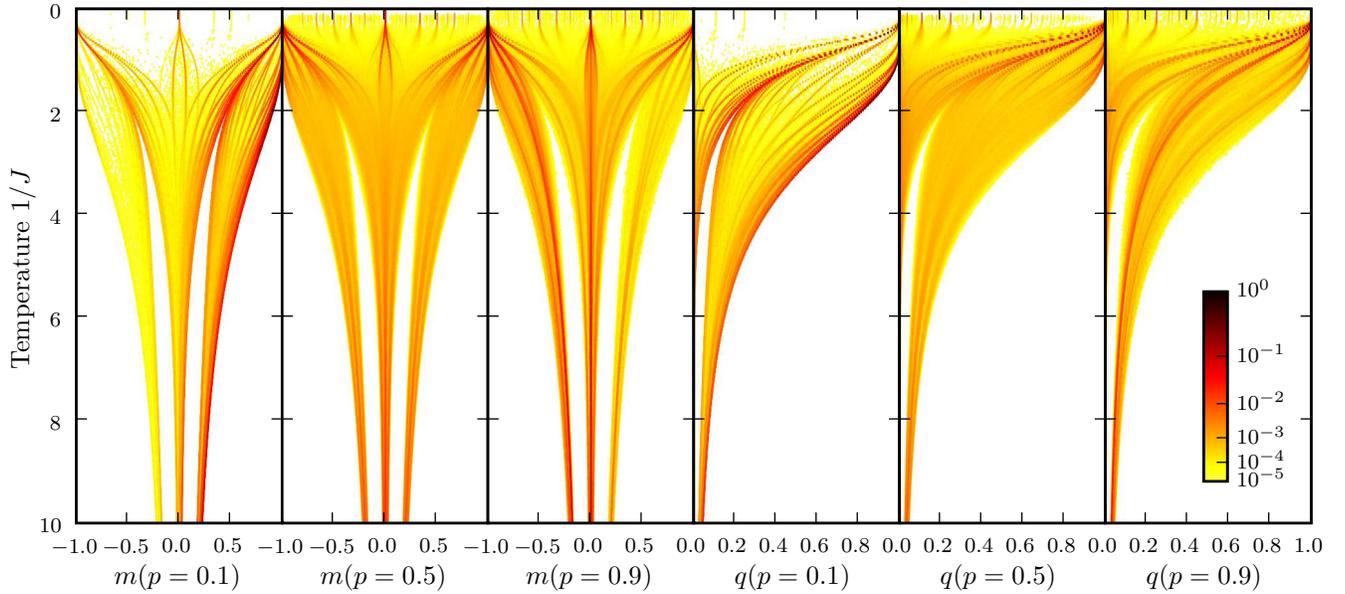}
\caption{(Color) Local order-parameter tulips: Probability
distributions of local magnetization (left panels) and spin-glass
(right panels) order parameters of the interior sites on an
Apollonian network with Ising spin-glass interactions, as a function
of temperature, for three different antiferromagnetic bond
concentrations $p$.}\label{fig:4}
\end{figure*}

Although their structure dates back to ancient Greek mathematics,
Apollonian networks~\cite{Andrade,DoyeMassen} have seen a recent
surge of interest as a simple and elegant model that incorporates
some of the key features identified in real-world networks: a
scale-free degree distribution, the small-world effect, and a high
clustering coefficient.  As such, they have become a versatile tool
for understanding the effects of complex topologies in interacting
systems: applications include percolation and epidemic
spreading~\cite{Andrade,Zhou}, magnetic systems~\cite{Andrade,
AndradeHerrmann}, mechanisms of network growth~\cite{Comellas},
avalanches in sandpile models~\cite{Vieira}, neural
networks~\cite{Pellegrini}, and even quantum behaviors like coherent
exciton transport~\cite{Xu} and correlated electron
models~\cite{Souza,Cardoso}.  The latter were inspired by the
development of synthetic, nanoscale, non-branched fractal
polymers~\cite{Newkome}, which raises the possibility that the
unique properties of scale-free structures like Apollonian networks
may be harnessed for technological applications.

In this study we focus on an intriguing aspect of these networks:
their ordering resilience in the presence of imposed quenched
disorder. For an Ising model with a variety of random-bond
distributions -- ferromagnetic/antiferromagnetic percolation and
spin-glass -- we find that ordered phases persist up to infinite
temperature, for every case except the pure antiferromagnetic system
(where the geometrical frustration of the network leads to
paramagnetism). The self-similar nature of the Apollonian network
allows the use of exact renormalization-group (RG) techniques to
calculate the phase diagram structure, even in the presence of
quenched randomness. While there have been many numerical RG studies
of spin glasses on fractal lattices, we have gone further,
developing an iterative procedure based on the local recursion
matrix that allows us to calculate exactly the complete distribution
of the local magnetization and spin-glass order parameters, for the
full range of bond probabilities and temperatures. The resulting
local-order diagrams (Fig.~\ref{fig:4}) show an intricate structure
as temperature is lowered, never before observed in such detail for
a disordered spin system.  Apollonian networks can be embedded in a
Euclidean plane without any edge
crossings~\cite{Andrade,DoyeMassen}.  This makes spin systems on
such networks potentially physically realizable, for example in
nanostructures formed from dense polydisperse packings of magnetic
grains~\cite{Andrade,AndradeHerrmann}.  Apollonian packings have
also been used in the study of smectic liquid crystals
\cite{Bidaux}.

\begin{figure}
\centering \includegraphics*[scale=1]{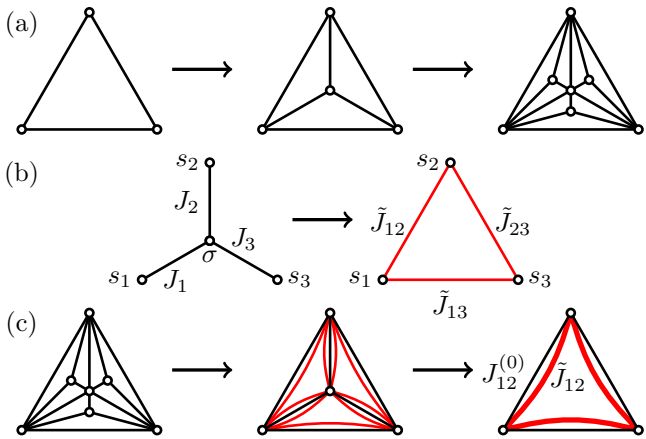}
\caption{(Color online) (a) Construction of an Apollonian network.
(b) Star-triangle transformation.  (c) Two successive RG
transformations of an Apollonian network.}\label{fig:1}
\end{figure}

\noindent {\em Network structure --} The construction of the
Apollonian network is depicted in Fig.~\ref{fig:1}(a): at each step,
a new site is added to the center of every triangle in the network,
and connected to the surrounding vertices.  In the limit of infinite
size, the geometrical characteristics of the network can be
summarized as follows~\cite{Andrade,DoyeMassen,Zhang3}:  $P(k)$
being the probability that a site has degree $k$, the cumulative
degree distribution is $P_\text{cum}(k) = \sum_{k^\prime =
k}^{\infty} P(k)\sim k^{1-\gamma}$ for large $k$, with the
scale-free exponent $\gamma = 1+\ln 3/\ln 2 \approx 2.585$.  Due to
the compact network structure, the average shortest-path length
$\bar{l}$ between any two points scales as in the small-world
effect, $\bar{l} \sim \ln N$, as shown in \cite{Zhang3} using the
exact recursive method of \cite{HinczewskiBerker}. As is typical in
small-world networks, the average clustering coefficient is large,
$C \approx 0.828$, measuring the ratio of the connections among the
nearest-neighbors of a site and the maximum possible number of such
connections $k(k+1)/2$, where $k$ is the degree of the site.  In all
these respects the topological properties of the Apollonian network
are comparable to those observed in empirical complex networks that
are simultaneously scale-free and small-world~\cite{BarabasiAlbert}.

We study an Ising Hamiltonian on the network, $-\beta {\cal H} =
\sum_{\langle ij \rangle} J_{ij} s_i s_j$, where $s_i=\pm 1$, $\beta
= 1/k_B T$, the sum $\langle ij \rangle$ is over nearest neighbors,
and the bond strengths $J_{ij}$ are distributed with a quenched
random probability distribution $P(J_{ij})$.  We consider two types
of distributions described by bond probability $p$~: (i) the
percolation case, where $P(J_{ij}) = p \delta(J_{ij}-J)+(1-p)
\delta(J_{ij})$, for both ferromagnetic (F) $J>0$ and
antiferromagnetic (AF) $J<0$ interactions; (ii) the $\pm J$ spin
glass (SG) case, where $P(J_{ij}) = p\delta(J_{ij}+J) +
(1-p)\delta(J_{ij}-J)$ with $J>0$.

\noindent {\em Exact renormalization-group transformation --} The
self-similar structure of the Apollonian network allows us to
formulate an exact RG transformation.  For $-\beta {\cal
  H}(\{J_{ij}\})$ the Hamiltonian for a particular configuration of
interactions on the $n$th-generation network (\textit{i.e.}, the
lattice after $n$ construction steps), the RG mapping yields a
Hamiltonian $-\beta^\prime {\cal H}^\prime(\{J^\prime_{ij}\})$ with
a renormalized set of interactions $\{J^\prime_{ij}\}$ on the
$(n-1)$th-generation network, preserving the partition function. The
mapping is carried out through a star-triangle transformation
\cite{Hilhorst}, tracing over the spins at sites added at the $n$th
step. This is shown for one plaquette in Fig.~\ref{fig:1}(b), with
the decimated spin labeled $\sigma$.  The trace over $\sigma$
produces interactions $\tilde J_{12}$, $\tilde J_{13}$, $\tilde
J_{23}$, between the edge sites of the triangle, which are functions
of the original interactions $J_1$, $J_2$, $J_3$ of the star,
$\tilde{J}_{12}=\frac{1}{4}ln
\left\{\frac{\cosh{\left(2J_1+2J_2\right)}+\cosh{\left(2J_3\right)}}{\cosh{\left(2J_1-2J_2\right)}+\cosh{\left(2J_3\right)}}\right\}$
and its cyclic permutations. In the context of the whole network,
the mapping works as shown in Fig.~\ref{fig:1}(c). The
$\tilde{J}_{ij}$ interactions (inner to each triangle) are added to
the original interactions $J_{ij}$ of the $(n-1)$th generation
network (in-between triangles), to give the renormalized
interactions $J^\prime_{ij}$.  Thus, in the bulk each original
interaction gets $\tilde{J}_{ij}$ contributions from its two
adjoining plaquettes.

In order to implement this RG transformation for the system in the
thermodynamic limit, we focus on the probability distribution of
triplets $Q(\{\tilde J_{ij}$, $\tilde J_{jk}$, $\tilde J_{ik}\})$
generated by the star-triangle transformation.  As we iterate the RG
mapping, this distribution $Q$ changes, and we can extract
thermodynamic information from the resulting flows.  To keep track
of $Q$ at each step, we adapt a numerical procedure developed by
Nobre~\cite{Nobre} for RG transformations of spin glasses on
hierarchical lattices.  This method has been shown to give
numerically accurate results for phase diagrams \cite{Ohzeki},
agreeing with more complicated binning techniques used to directly
evaluate the RG flows of interaction distributions
\cite{HinczewskiBerker}.  We represent the distribution $Q$ by a
pool of large size $M$, where each element in the pool consists of a
triplet of real numbers.  To generate the initial pool $Q^{(1)}$, we
repeat the following $M$ times: (i) choose three random numbers
$J_1, J_2, J_3$ with the probability $P(J)$; (ii) perform the
star-triangle transformation of Fig.~\ref{fig:1}(b), yielding a
triplet $\{\tilde J_{12},\tilde J_{23},\tilde J_{13}\}$ which is
placed in the pool.  Each subsequent RG transformation creates a new
pool $Q^{(i)}$ from the previous pool $Q^{(i-1)}$ in the following
manner, again repeating the same procedure $M$ times to preserve the
size of the pool: (i) randomly choose three triplets from
$Q^{(i-1)}$; (ii) randomly arrange these three triplets like the
three triangles in the second step of Fig.~\ref{fig:1}(c), together
with the three middle bonds, chosen randomly with probability
$P(J)$; (iii) decimate the center spin to yield a renormalized
triplet, namely the inner bonds in the third step of
Fig.~\ref{fig:1}(c)), which is placed in pool $Q^{(i)}$.  As $M \to
\infty$, the pools mimic the exact renormalized distributions of
triplets in the thermodynamic limit.  For the present work, we found
that $M = 10^6$ was sufficiently large to make finite-ensemble
effects negligible.  From the behaviors of the $Q^{(i)}$ in the
limit of large $i$, we can identify the phase structure of the
system. Specifically, looking at the average $\bar{J}^{(i)}$ and
standard deviation $\sigma_J^{(i)}$ of the $3M$ bond strengths in
pool $Q^{(i)}$, we can distinguish three limiting behaviors as $i
\to \infty$: (i) a ferromagnetic (F) sink, where $\bar{J}^{(i)} \to
\infty$, $\sigma_J^{(i)} \to \infty$, $\sigma_J^{(i)}/\bar{J}^{(i)}
\to 0$; (ii) a spin-glass (SG) sink, where $\bar{J}^{(i)} \to
\infty$, $\sigma_J^{(i)} \to \infty$, $\bar{J}^{(i)}/\sigma_J^{(i)}
\to 0$; (iii) a paramagnetic (P) sink, where $\bar{J}^{(i)} \to 0$,
$\sigma_J^{(i)} \to 0$. Furthermore, the RG evolution of fraction of
frustrated triangles is different in the ferromagnetic and
spin-glass phases, respectively going to 0 and 0.5 in the
ferromagnetic and spin-glass phases.

\begin{figure}
\includegraphics*[scale=1]{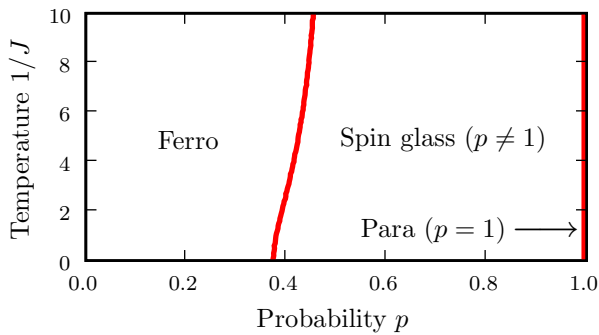}
\caption{(Color online) Phase diagram of the Ising spin glass on an
Apollonian network, in temperature $1/J$ versus antiferromagnetic
bond concentration $p$. The boundary between the ferromagnetic and
spin-glass phases is first order. The paramagnetic phase appears
with a first-order phase transition at $p = 1$.}\label{fig:2}
\end{figure}

\noindent {\em Calculation of local magnetizations and local SG
order parameters --} Moreover, the numerical procedure described
above is not limited to just the distribution of renormalized
interactions $Q$. It can be extended to determine additional
thermodynamic details, in particular the distribution of local
magnetizations and local SG order parameters. Let us consider the
magnetization $m_\sigma$ at a site $\sigma$ in the original lattice.
For simplicity, let $\sigma$ be one of the sites generated at the
last construction step.  We shall denote these as ``interior
sites'', and they constitute 2/3 of the total lattice in the limit
of large $n$. Adding a local magnetic field $H_\sigma$, we can write
$m_\sigma =
\partial \ln Z/\partial H_\sigma|_{H_\sigma=0}$, where $Z$ is the
partition function.  The star-triangle transformation with
$H_\sigma$ produces additional interactions in the renormalized
triangle on the right side of Fig.~\ref{fig:1}(b): three local
fields $H_1 s_1, H_2 s_2, H_3 s_3$, and a three-site term $K s_1 s_2
s_3$, where $H_i$ and $K$ are functions of the $J_{ij}$ and
$H_\sigma$. With these interactions, the RG mapping is closed upon
further iteration, so there will be a set of parameters ${\bf
K}^{(i)} \equiv \{H_1^{(i)},H_2^{(i)},H_3^{(i)},K^{(i)}\}$ after the
$i$th RG step associated with the triangle that originally contained
spin $\sigma$. Using the chain rule, $m_\sigma$ can be expressed
\cite{McKay, Yesilleten} in terms of local recursion matrices $T$
over the $i$ steps,
\begin{equation}\label{eq:1}
m_\sigma = {{\boldsymbol{\mu}^{(i)}}^T} T^{(i)}T^{(i-1)}\cdots T^{(2)} {\bf V}^{(1)},
\end{equation}
where $\mu^{(i)}_\alpha = \partial \ln Z/\partial K^{(i)}_\alpha$,
$T^{(i)}_{\beta\alpha} = \partial K^{(i)}_\beta / \partial
K^{(i-1)}_\alpha$, $V^{(1)}_\alpha = \partial
K_\alpha^{(1)}/\partial H_\sigma$, and $K^{(i)}_\alpha$ are the
components of ${\bf K}^{(i)}$. All these quantities are evaluated in
the RG subspace with initial condition $H_\sigma =0$ (\textit{i.e.},
where $K^{(i)}_\alpha = 0$ for all $i$), which makes them functions
only of the $J_{ij}$ configuration at the previous step.  In the
thermodynamic limit, as a corner boundary condition, we calculate
$\boldsymbol{\mu}^{(i)}$ over the up-magnetized configurations of
the three original corner spins of the network. As the RG transform
is iterated, each triplet in the pool $Q^{(1)}$ has a corresponding
vector $T^{(i)}\cdots T^{(2)}{\bf V}^{(1)}$, which can be contracted
with $\boldsymbol{\mu}^{(i)}$ calculated from $Q^{(i)}$ to obtain a
pool of $m_\sigma$ using Eq.~\eqref{eq:1}. For sufficiently large
$i$ and $M$, the resulting pool converges to the exact distribution
of local magnetizations in the thermodynamic limit.  The averages of
$m_\sigma$ and $q_\sigma \equiv m_\sigma^2$ over this distribution
respectively yield the magnetization $m$ and SG order parameter $q$
for the interior sites.

\begin{figure}
\includegraphics*[scale=1]{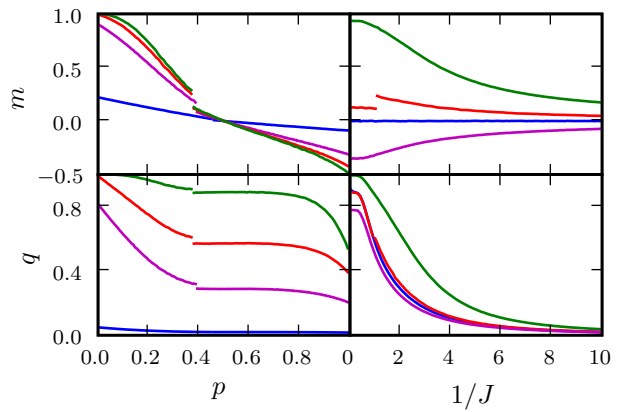}
\caption{(Color online) Magnetization $m$ and spin-glass order
parameter $q$ for the Ising spin glass on an Apollonian network.  In
the left panels, the curves, consecutively leftmost from the top,
are for constant temperature $1/J = 0.2, 1.0, 2.0, 10.0$.  In the
right panels, the curves, consecutively from the top, are for
antiferromagnetic bond concentration $p = 0.10, 0.38, 0.50,
0.90$.}\label{fig:3}
\end{figure}

\noindent {\em Results --} We focus first on the ferromagnetic (F)
and antiferromagnetic (AF) percolation cases.  For F percolation the
system is ferromagnetically ordered at all finite temperatures for
any $p > 0$, directly related to the existence of a giant connected
component in the network at all nonzero bond
probabilities~\cite{Andrade}.  In contrast, one might expect
macroscopic order in the system to be inhibited in the AF case,
since AF bonds are frustrated on the triangular plaquettes in the
network. Indeed, in the case of $p=1$, namely for a pure AF system,
frustration leads to a paramagnetic phase at all temperatures.
However, as soon as even a tiny fraction of AF bonds is removed,
namely for any $0 < p < 1$, we find an SG phase {\em at all
temperatures}, an interesting example of a glassy phase which is
completely impervious to thermal excitations.  The SG phase appears
even when weaker forms of disorder are added to the pure AF system,
such as simply attenuating a fraction of the bonds. Consider the
range of models described by the bond distribution $P(J_{ij}) = p
\delta(J_{ij}-J)+(1-p) \delta(J_{ij}-cJ)$ where $J < 0$ and $0 \le c
< 1$.  Here $c=0$ corresponds to the AF percolation case described
above, but it turns out that any $c < 1$ gives the same phase
diagram structure: a SG phase of infinite extent for all $0 < p < 1$
and paramagnetism at $p=1$.

We now turn to the spin-glass system: The system composed of
antiferromagnetic bonds, under infinitesimal doping by ferromagnetic
bonds, produces a spin-glass phase via a jump, namely a first-order
phase transition at $p=1$. For a sufficient quantity of
ferromagnetic doping, a first-order transition occurs to the
ferromagnetic phase, as can seen in the phase diagram in
Fig.~\ref{fig:2}. On both sides of the transition line the ordered
phases persist to infinite temperature $1/J$, and the boundary
itself asymptotically approaches a vertical line at $p = 0.5$ as
$1/J \to \infty$.  The first-order nature of the
ferromagnetic-spinglass phase transition, to our knowledge not seen
in other systems, is evident from the magnetization and SG order
parameter plotted in Fig.~\ref{fig:3}, which indeed show
discontinuities crossing the boundary. (At the highest temperature
depicted, $1/J = 10.0$, the discontinuities exist but are too small
to be seen on the scale of the figure.) Fig.~\ref{fig:3} also
reveals a curious aspect of spin-glass order on the Apollonian
network: unlike a conventional spin-glass phase, the magnetization
$m$ is generally nonzero. With the above-mentioned boundary
condition on the three corner spins, $m \gtrless 0$ for $p \lessgtr
0.5$. The negative $m$ at large $p$ is understood from the influence
of the corner spins, which except for the central spin have the
highest degree in the network. In an environment with mostly
antiferromagnetic bonds, an up orientation for the corner spins will
yield a negative magnetization of the interior sites. This ability
of the most connected spins to determine the sign of the
magnetization may be a general feature of scale-free networks, and
has been seen in the Barab\'asi-Albert model~\cite{Aleksiejuk}.

The evolution of the system under disorder is obtained in
microscopic detail by the calculation of the local magnetizations
and local SG order parameters, as described above.  The resulting
full distributions of the local magnetizations and SG order
parameters are given in Fig.~\ref{fig:4}. To produce these graphs,
the $m_\sigma$ and $q_\sigma$ pools were coarse-grained using a
binning procedure, and the normalized heights of the resulting
histograms color-coded. For clarity, histograms smaller than
$10^{-5}$ are not shown.  The distributions exhibit a distinctive
tuliplike shape, developing a rich structure as the system is
cooled, spreading from narrowly localized peaks at high temperatures
into complex bands of smaller peaks over the whole range at
intermediate temperatures. These bands in turn converge toward the
expected sharply defined values at low temperatures, with local
magnetization peaked around 1, 0, and -1, and the SG order parameter
around 0 and 1. The asymmetry leading to negative magnetization $m$
for $p > 0.5$ is evident in comparing the $p=0.5$ and $p=0.9$ local
magnetization plots.  The former is entirely symmetric between
negative and positive peaks, while in the latter the predominance of
antiferromagnetic bonds leads to the bands of negative peaks
becoming more prominent.

In conclusion, we have shown that ferromagnetic phases and,
moreover, spin-glass phases on Apollonian networks exhibit a
remarkable robustness, with an infinite critical temperature for any
amount of disorder. In fact, order persists to infinite temperature
even when almost all of the bonds in the system are removed and even
when almost all of the bonds in the system are frustrated, namely in
the $p$ infinitesimally greater than zero and in the $p$
infinitesimally less than one regimes of the percolation and
spin-glass problems, respectively. This property should have
consequences for actual applications on networks. For example,
interacting objects arranged on Apollonian nanostructures would be
able to maintain cooperative behavior over a broad range of
temperatures and intrinsic disorder. Our local renormalization-group
theory method yields, in the network with frozen disorder, the exact
local order parameters, up to now only calculated approximately by
mean-field theory.  The resulting local magnetizations and local
spin-glass order parameters do not yield just a distribution of
values, as would be most simply expected in systems with frozen
disorder, but also unexpectedly distinctive tulip structures with
stalks, leaves, and veins, evolving under temperature.  It is
clearly unlikely that the new and intriguing tulip structures, with
stalks, leaves, and veins, in the microscopics are limited to
Apollonian networks, but more likely will appear, perhaps in varying
topologies, in diverse small-world systems.  Our locally
discriminating renormalization-group technique, in yielding such
detailed local results, should be of interest for the positional
distribution of order in systems with inhomogeneities, be it due to
quenched impurities or surfaces, \textit{etc}.

\noindent {\em Acknowledgments -} This research was supported by the
Scientific and Technological Research Council of Turkey
(T\"UB\.ITAK) and by the Academy of Sciences of Turkey.

\end{document}